\documentclass[reprint, superscriptaddress,  amsmath, amssymb, aps, prl]{revtex4-2}
\usepackage{dcolumn, braket, graphicx, xcolor, bm, lipsum}

\begin{document}

\title{Dirac half-semimetallicity and antiferromagnetism \\ in graphene nanoribbon/hexagonal boron nitride heterojunctions}

\author{Nikita~V.~Tepliakov}
\email{n.tepliakov20@imperial.ac.uk}
\affiliation{Departments of Materials and Physics, Imperial College London, London SW7 2AZ, United Kingdom}
\affiliation{The Thomas Young Centre for Theory and Simulation of Materials, Imperial College London, London SW7 2AZ, United Kingdom}

\author{Ruize~Ma}
\affiliation{Departments of Materials and Physics, Imperial College London, London SW7 2AZ, United Kingdom}
\affiliation{The Thomas Young Centre for Theory and Simulation of Materials, Imperial College London, London SW7 2AZ, United Kingdom}
\affiliation{Department of Physics, University of Oxford,  Oxford OX1 2JD, United Kingdom}

\author{Johannes~Lischner}
\affiliation{Departments of Materials and Physics, Imperial College London, London SW7 2AZ, United Kingdom}
\affiliation{The Thomas Young Centre for Theory and Simulation of Materials, Imperial College London, London SW7 2AZ, United Kingdom}

\author{Efthimios~Kaxiras}
\affiliation{Department of Physics, Harvard University, Cambridge, MA 02138, United States}
\affiliation{School of Engineering and Applied Sciences, Harvard University, Cambridge, MA 02138, United States}

\author{Arash~A.~Mostofi}
\affiliation{Departments of Materials and Physics, Imperial College London, London SW7 2AZ, United Kingdom}
\affiliation{The Thomas Young Centre for Theory and Simulation of Materials, Imperial College London, London SW7 2AZ, United Kingdom}

\author{Michele~Pizzochero}
\email{mpizzochero@g.harvard.edu}
\affiliation{School of Engineering and Applied Sciences, Harvard University, Cambridge, MA 02138, United States}

%\date{}

\begin{abstract} 
Half-metals have been envisioned as active components in spintronic devices by virtue of their completely spin-polarized electrical currents. Actual  materials hosting half-metallic phases, however, remain scarce. Here, we predict that recently fabricated heterojunctions of zigzag nanoribbons embedded in two-dimensional hexagonal boron nitride are half-semimetallic, featuring fully spin-polarized Dirac points at the Fermi level. The half-semimetallicity originates from the  transfer of charges from hexagonal boron nitride to the embedded graphene nanoribbon. These charges give rise to opposite energy shifts of the states residing at the two edges while preserving their intrinsic antiferromagnetic exchange coupling. Upon doping, an antiferromagnetic-to-ferrimagnetic phase transition occurs in these heterojunctions, with the sign of the excess charge controlling the spatial localization of the net magnetic moments. Our findings demonstrate that such heterojunctions realize tunable one-dimensional conducting channels of spin-polarized Dirac fermions that are seamlessly integrated into a two-dimensional insulator, thus holding promise for the development of carbon-based spintronics.
 \end{abstract}

\maketitle

Graphene nanoribbons (GNRs) --- few nanometer-wide strips of $sp^2$-bonded carbon atoms  --- are promising components for next-generation nanoscale electronics \cite{Yazyev2013, Wang2021a} because of their sizable energy gap \cite{Son2006a, Han2007, Chen2013, Tepliakov2023}, superior charge transport \cite{Baringhaus2014}, and facile integration into short-channel field-effect transistors \cite{Xiaolin2008, Bennett2013, Llinas2017,  BorinBarin2019}. GNRs can be fabricated in an atomically precise fashion \cite{Cai2010a}, leading to a rich spectrum of edge geometries \cite{Zongping2020, Yano2020} and electronic phases \cite{Cao2017, Groning2018, Rizzo2018, Sun2020b, Arnold2022}. Of particular interest are zigzag graphene nanoribbons (ZGNRs) \cite{Ruffieux2016}, owing to their magnetically ordered ground state \cite{Fujita1996, Yazyev2010, Blackwell2021} that can be engineered through charge doping \cite{Jung2010}, electric fields \cite{Son2006b, Kan2007}, lattice deformations \cite{Hu2012, Zhang2017}, or chemical functionalization of the edges \cite{Kan2008, Pizzochero2021, Pizzochero2022}. The combination of controllable $\pi$-electron magnetism with a long spin coherence time at room temperature \cite{Slota2018, Magda2014} renders ZGNRs suitable building blocks for spin logic operations \cite{Yazyev2008}.

In contrast to the wide application of graphene \cite{Liu2013a, Liu2014b, Gui2016c, Chen2017d} and other atomically thin crystals \cite{Huang2014e, Lin2014f}, the integration of GNRs into complex junctions remains relatively unexplored. It is limited to in-plane homojunctions \cite{Cai2010a} resulting from the lateral connection of graphene nanoribbons with distinct widths \cite{Chen2015a, Jacobse2017, Wang2017, Pizzochero2020a}, crystallographic orientations \cite{Cai2010a, Dienel2015}, or heteroatom incorporations \cite{Cai2014, Nguyen2017}. Recently, progress has been made through the realization of heterojunctions consisting of one-dimensional graphene nanoribbons of target chirality, including ZGNRs, embedded into a continuous two-dimensional hexagonal boron nitride ($h$BN) sheet \cite{Chen2017, Wang2021b}. The fabrication of such ZGNR/$h$BN heterojunctions is accomplished in two steps \cite{Wang2021b}. First, nanotrenches along the zigzag direction are carved in $h$BN through nickel nanoparticle-catalyzed cutting. Second, these trenches are filled with ZGNRs obtained through chemical vapor deposition. The integrated growth of ZGNRs into $h$BN offers an effective route to achieve ultrathin, large-scale nanocircuitry. Yet, the prospect of employing these heterojunctions in real devices requires a detailed understanding of their functionalities. Earlier theoretical works have shown that crystallographically aligned \cite{Leon2019} or misaligned  superlattices \cite{Zeng2016} and interfaces \cite{liu2011half} of graphene and $h$BN nanoribbons may feature half-metallicity \cite{ding2009electronic,Pruneda2010}, which is attributed to the electronic states localized at the boundary between the constituent materials. Though instructive, the adopted computational models are however insufficient to provide a realistic description of the recently fabricated graphene nanoribbons embedded in $h$BN.

\begin{figure*}[t!]
    \centering
\includegraphics[width=1.75\columnwidth]{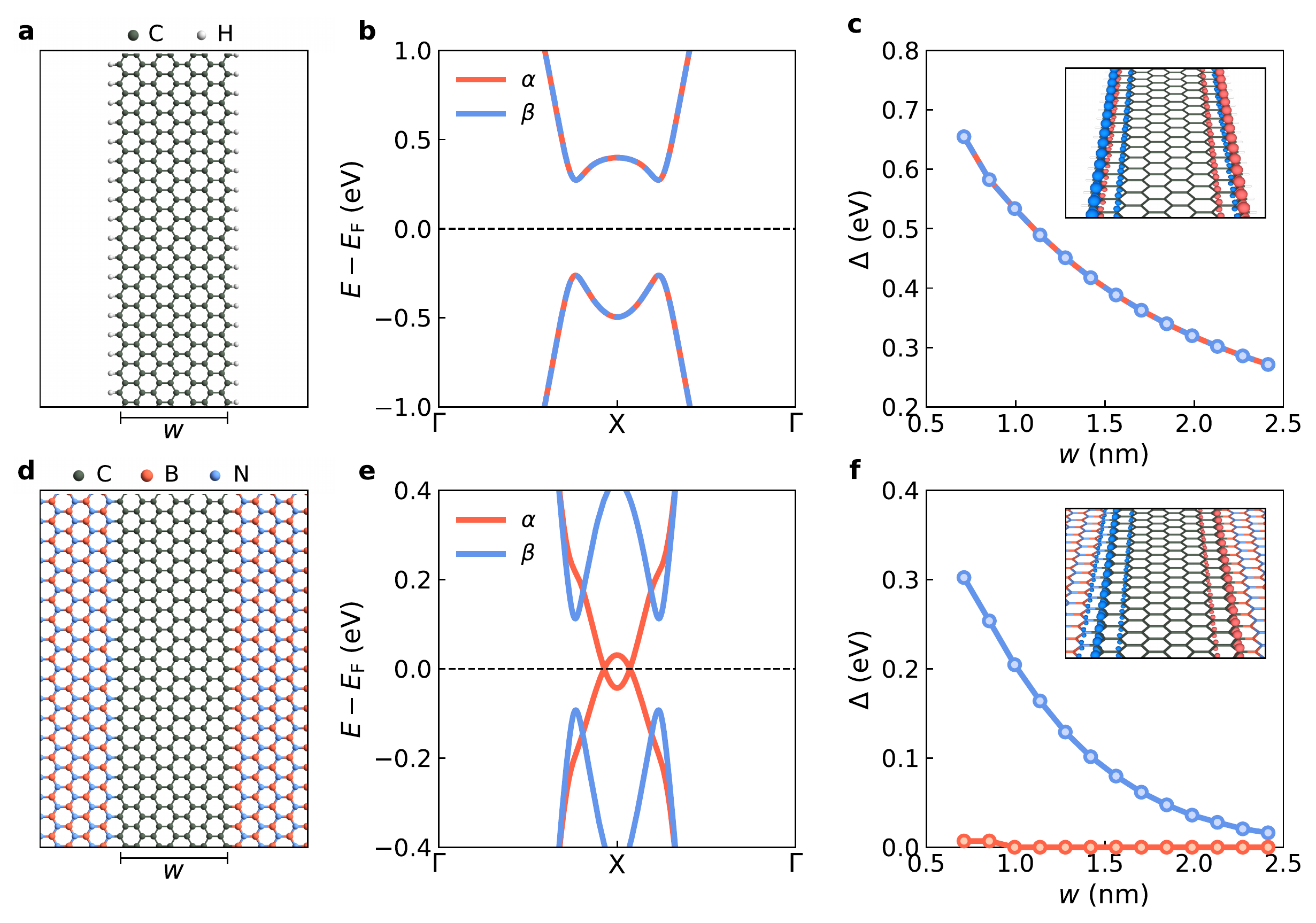}
    \caption{\textbf{Dirac half-semimetallicity in ZGNR/\textit{h}BN heterojunctions.}
    (a) Atomic structure of a hydrogen-terminated ZGNR of width $w = 1.4$~nm, which is periodic and infinite in the vertical direction. (b) Band structure of the hydrogen-terminated ZGNR shown in panel (a); red and blue lines denote $\alpha$- and $\beta$-spin bands; energies are referenced to the Fermi level ($E$\textsubscript{F}). (c) Band gaps ($\Delta$) of hydrogen-terminated ZGNRs as a function of the width ($w$). (d) Atomic structure of a ZGNR/$h$BN heterojunction comprising a ZGNR of width $w = 1.4$~nm embedded in a two-dimensional $h$BN matrix. (e) Band structure of the ZGNR/$h$BN heterojunction shown in panel (d). (f) Band gaps ($\Delta$) of ZGNR/$h$BN heterojunctions as a function of the ZGNR width ($w$). The insets in panels (c) and (f) show the spin density of a hydrogen-terminated ZGNR and a ZGNR/$h$BN heterojunction, respectively, for $w = 1.4$~nm; red and blue contours represent $\alpha$- and $\beta$-spin density.}
    \label{Fig1}
\end{figure*}

In this Letter, we investigate zigzag graphene nanoribbons embedded in a two-dimensional sheet of hexagonal boron nitride on the basis of \emph{ab initio} and mean-field Hubbard Hamiltonian calculations. We show that such ZGNR/$h$BN heterojunctions are half-semimetallic, as the semiconducting behavior in one spin orientation is accompanied by a Dirac semimetallic behavior in the other, while preserving the intrinsic antiferromagnetism of ZGNRs. This unconventional combination of spin-split, Dirac half-semimetallicity and antiferromagnetism stems from the charge transfer at the interfaces between the magnetic edges of the nanoribbon and the hexagonal boron nitride. Charge doping can further modulate the electronic structure of the heterojunctions by enforcing a ferrimagnetic ground state and enabling precise control over the spatial localization of magnetic moments. Thus, these nanoarchitectures  realize one-dimensional conducting channels of spin-polarized Dirac fermions embedded within a two-dimensional insulator, opening new avenues for the exploration of spintronic devices based on graphene.

Our \emph{ab initio} calculations are based on semilocal density-functional theory (DFT). The computational details are provided in the Supporting Information (Supporting Notes S1 and S2). For comparison, we begin by recalling the electronic properties of hydrogen-terminated zigzag graphene nanoribbons, whose atomic structure for a representative width of 1.4 nm is given in Figure~\ref{Fig1}(a).  ZGNRs possess width-independent $\pi$-electron magnetic moments of $0.27\,\mu_\mathrm{B}$ per unit cell that are localized at the edge carbon atoms. The magnetic moments align parallel along each edge of the nanoribbon and antiparallel across opposite edges. The inter-edge antiferromagnetic exchange coupling, obtained as the difference in energy between parallel and antiparallel spin orientations, is 3.64 meV. In line with Lieb's theorem for the repulsive Hubbard model on a bipartite lattice at half filling \cite{Lieb1989}, this leads to a spin-zero ground state, as evident from the spin-density pattern shown in the inset of Figure~\ref{Fig1}(c). The antiferromagnetic ordering of the edge-localized magnetic moments in ZGNRs results in the spin-degenerate band structure shown in Figure~\ref{Fig1}(b), which exhibits a direct band gap. As reported in Figure~\ref{Fig1}(c), the magnitude of the energy gap decreases monotonically as the width of the nanoribbon increases due to the weakening of confinement effects. Hence, ZGNRs are semiconducting and antiferromagnetic regardless of their width.

\begin{figure*}[t!]
    \centering
\includegraphics[width=1.85\columnwidth]{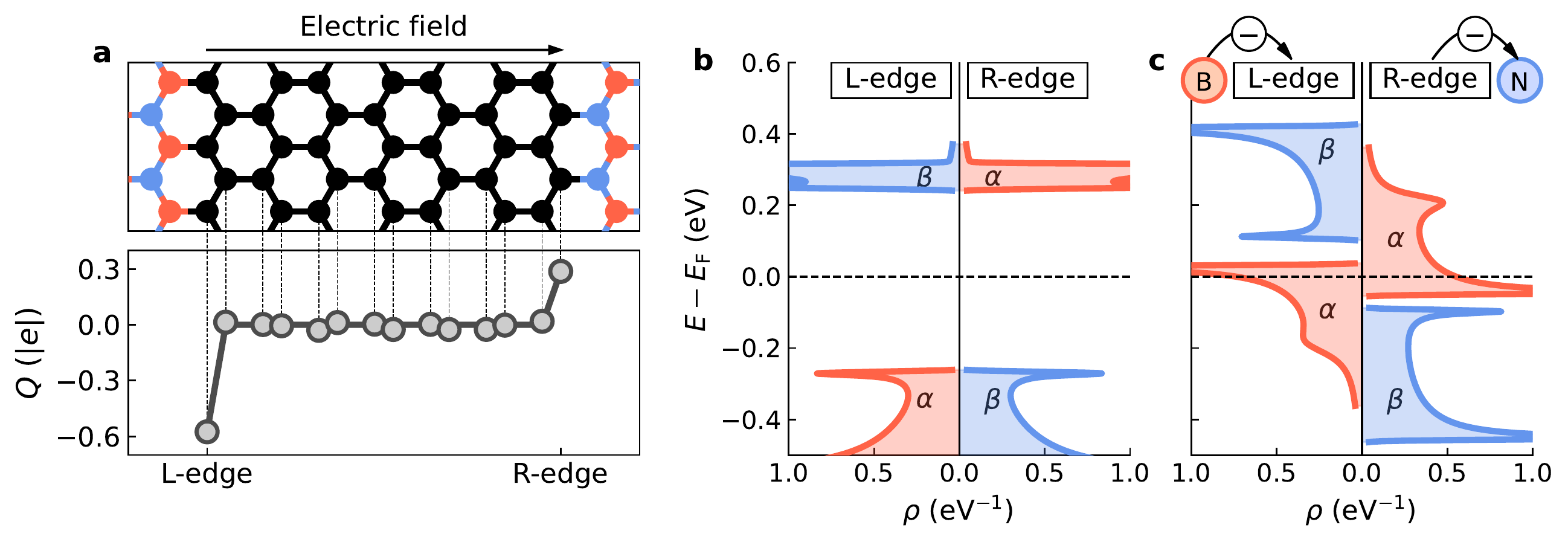}
    \caption{
    \textbf{Mechanism underlying the half-semimetallicity in ZGNR/\textit{h}BN heterojunctions.}
    (a) Atomic charges ($Q$) at each carbon atom across the ZGNR/$h$BN heterojunction with $w=1.4$~nm; in the upper panel, blue, red, and black spheres indicate nitrogen, boron, and carbon atoms, respectively. (b) Density of states ($\rho$) of hydrogen-terminated ZGNRs projected onto the carbon atoms at the left (L-) and right (R-) edge of the nanoribbon. (c) Projected density of states of the ZGNR/$h$BN heterojunction, where electrons transferred from boron-to-carbon (carbon-to-nitrogen) atoms at the interface between $h$BN and the L-edge (R-edge) of ZGNR induce an upward (downward) energy shift of each pair of edge states.}
    %\textbf{b}, Schematics of the electronic band diagram of hydrogen-terminated ZGNRs, where the left (L-) and right (R-) edge provides a pair of degenerate bands of opposite spin. \textbf{c}, Schematics of the electronic band diagram of ZGNR/$h$BN heterojunctions, where the  charge transferred from boron-to-carbon (carbon-to-nitrogen) atoms at the interface between $h$BN and the L-edge (R-edge) of ZGNR induces an upward (downward) energy shift of each pair of edge states.}
    \label{Fig2}
\end{figure*}

We next consider heterojunctions of zigzag graphene nanoribbons of the same width of 1.4~nm embedded in hexagonal boron nitride. Because of the structural topology of the ZGNR, its incorporation into $h$BN enforces a connectivity pattern such that the carbon atoms along opposite edges of the nanoribbon are covalently bonded to boron and nitrogen atoms, as seen in Figure~\ref{Fig1}(d). Importantly, the $h$BN matrix does not modify the spin density distribution, where edge-localized magnetic moments retain an antiparallel alignment that leads to a zero net magnetization, as shown in the inset of Figure\ref{Fig1}(f). Compared to hydrogen-terminated ZGNR, the magnetic moment is lowered to $0.19\,\mu_\mathrm{B}$ per unit cell, with the antiferromagnetic exchange coupling reduced to 2.41 meV. The electronic band structure of the ZGNR/$h$BN heterojunction is qualitatively different from that of hydrogen-terminated ZGNR. In particular, the ZGNR/$h$BN heterojunction exhibits the spin-split band structure shown in Figure~\ref{Fig1}(e), where the $\alpha$-spin electrons are semimetallic and $\beta$-spin are semiconducting. The crossing of the $\alpha$-spin energy bands forms Dirac points at the Fermi level, with a corresponding Fermi velocity of $1.2 \times 10^5$~m/s, approximately one order of magnitude lower than that of monolayer graphene~\cite{Geim2013}. This is the hallmark of Dirac half-semimetallicity \cite{deGroot1983} which, as shown in Figure~\ref{Fig1}(f), is insensitive to the width of the nanoribbon embedded in $h$BN.  Hence,  ZGNR/$h$BN heterojunctions host an unconventional Dirac half-semimetallic antiferromagnetic phase \cite{Xiao2012}, where a spin-split band structure, typical of ferromagnets, coexists with fully compensated magnetization, typical of antiferromagnets.

The emergence of the half-semimetallic phase in the ZGNR/$h$BN heterojunctions traces back to charge transfer occurring at the two interfaces between the constituent materials. In Figure~\ref{Fig2}(a), we show the atomic charges residing at each carbon atom in the ZGNR/$h$BN heterojunction with $w=1.4$~nm, as obtained from an analysis based on Bader's theory of atoms in molecules~\cite{Tang2009}.  Carbon atoms on the edges of the nanoribbon are negatively charged when bonded to boron atoms and positively charged when bonded to nitrogen atoms. This is a consequence of the difference in electronegativity between these elements \cite{Pizzochero2015}. Because of the formation of an array of point-like charges of opposite polarity at the spatially-separated edges, a built-in electric field develops in the in-plane direction perpendicular to the periodic direction of the heterojunction. In Figure~\ref{Fig2}(b) and \ref{Fig2}(c), we show the electronic density of states projected onto the edge carbon atoms. The total density of states is given in the Supporting Information (Supporting Note S3). The half-semimetallicity arises from the energy shifts of opposite signs caused by the built-in electric field, which lifts the degeneracy of the localized edge states. Specifically, the negatively charged edge of the ZGNR shifts the $\alpha$-spin valence band and $\beta$-spin conduction band towards higher energies, whereas the positively charged edge shifts the $\alpha$-spin conduction band and $\beta$-spin valence band towards lower energies. Consequently, the energy gap of the $\alpha$-spin closes. This is qualitatively analogous to previously investigated external field effects in hydrogen-terminated armchair \cite{Raza2008, Roslyak2010, Pizzochero2021a} and zigzag graphene nanoribbons \cite{Son2006b}, but here the behavior is driven by intrinsic charge transfer resulting from embedding the ZGNR in a two-dimensional $h$BN matrix. Our results are consistent with previous calculations performed on superlattices consisting of alternating graphene and hexagonal boron nitride nanoribbons \cite{ding2009electronic,Pruneda2010, Zeng2016, Leon2019} and lend support to the experimental observation of a finite electronic conductance at the Fermi level in these heterojunctions \cite{Wang2021b}. 

\begin{figure}[t!]
    \centering
    \includegraphics[width=0.85\columnwidth]{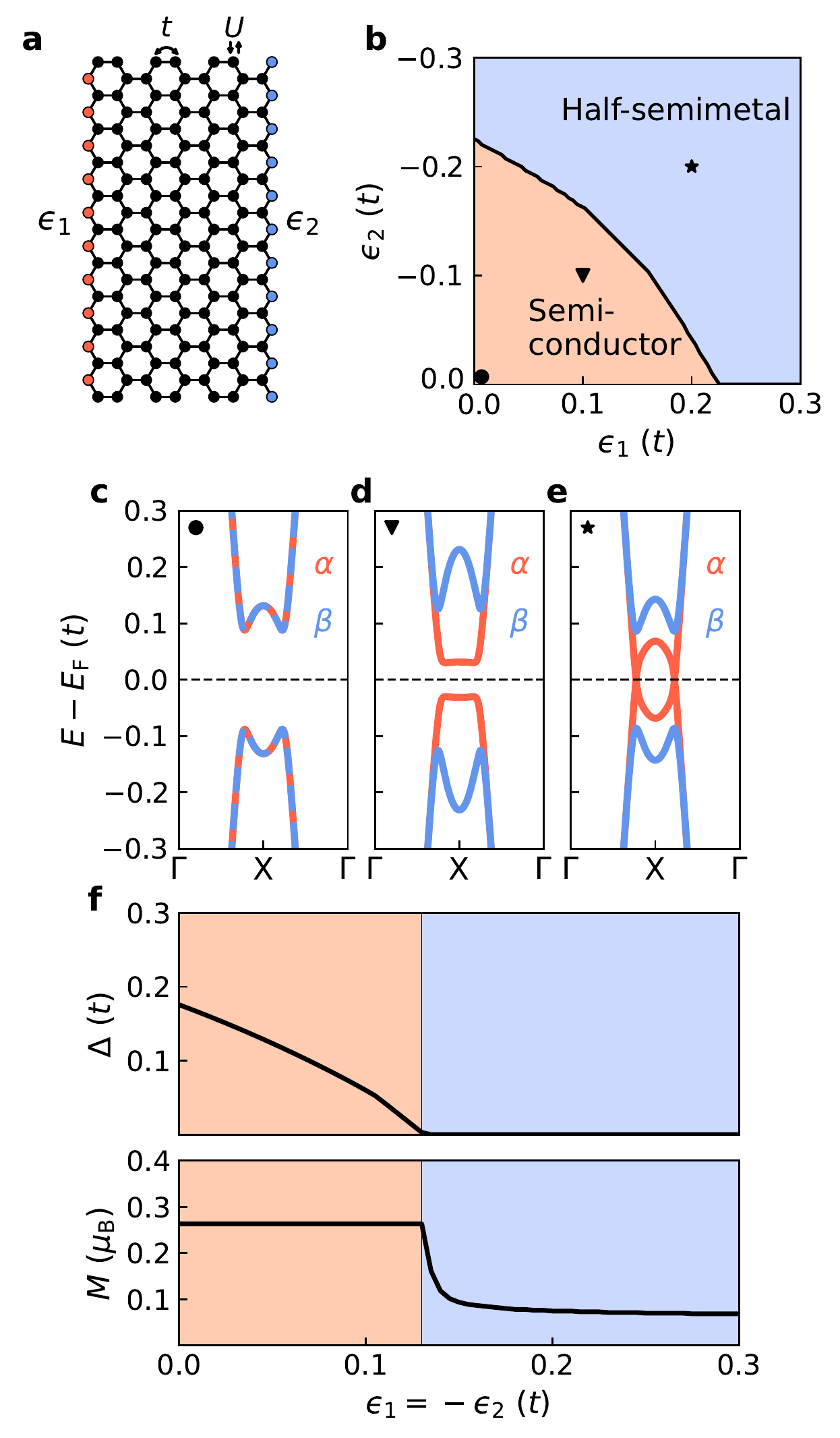}
    \caption{\textbf{Interplay between magnetic edge states and charge transfer in ZGNR/\textit{h}BN heterojunctions.}
    (a) Schematic representation of the mean-field Hubbard model used to describe a ZGNR embedded in $h$BN; black, red, and blue dots denote lattice sites with on-site potentials of $\epsilon = 0$, $\epsilon = \epsilon_1$, and $\epsilon = -\epsilon_2$ respectively; $t$ is the nearest-neighbor hopping amplitude and $U$ is the on-site Coulomb repulsion energy of a pair $p_z$-electrons residing at the same lattice site. (b) Electronic phase diagram of ZGNR as a function of on-site potentials ($\epsilon_1$ and $\epsilon_2$) on the edges; red and blue shaded areas correspond to semiconducting and half-semimetallic phases, respectively. Representative band structures of ZGNR at (c) $\epsilon_1 = \epsilon_2 = 0$, (d) $\epsilon_1 = -\epsilon_2 = 0.1t$, and (e) $\epsilon_1 = -\epsilon_2 = 0.2t$; red and blue lines denote $\alpha$-spin and $\beta$-spin energy bands, respectively. (f) Energy gaps ($\Delta$) and magnetic moments per carbon atom on the edge ($M$) as a function of on-site potentials $\epsilon_1 = -\epsilon_2$. In all the panels, the width of the nanoribbon is $w=1.4$~nm.}
    \label{Fig3}
\end{figure}

To gain insight into the origin of the half-semimetallicity in ZGNR/$h$BN heterojunctions and further elucidate the role of charge transfer in the magnetic states at the zigzag edges, we develop a simple and intuitive model restricted to the $\pi$-electron network of ZGNRs. This model is described by the following mean-field Hubbard model (MFHM) Hamiltonian  \cite{Yazyev2008},
\begin{equation}\label{hubbard}
\begin{split}
\hat{\cal H} = -t \sum_{\braket{i,j}, \sigma} \left[\hat c_{i\sigma}^{\dagger} \hat c_{j\sigma} + \textnormal{h.c.}\right] + \sum_{i,\sigma} \epsilon_i \hat c_{i\sigma}^{\dagger} \hat c_{i\sigma} +  \\ U \sum_i \left[\hat n_{i\uparrow}\langle\hat n_{i\downarrow}\rangle + \langle\hat n_{i\uparrow}\rangle\hat n_{i\downarrow}\right],
\end{split}
\end{equation}
where $\hat c_{i\sigma}$ and $\hat c_{i\sigma}^{\dagger}$ are the annnihilation and creation operators, respectively, for a $p_z$-electron with spin $\sigma$ at lattice site $i$ ($\mathrm{h.c.}$ denotes the Hermitian conjugate), $t$ is the hopping amplitude between nearest-neighboring lattice sites $i$ and $j$, $\epsilon_i$ is the on-site potential at lattice site $i$, $\hat n_{i\sigma} = \hat c_{i\sigma}^{\dagger} \hat c_{i\sigma}$ is the spin density at lattice site $i$, and $U$ is the strength of the on-site Coulomb repulsion between a pair of $p_z$-electrons residing at the same lattice site. We use $U = t$ in line with earlier \emph{ab initio} calculations \cite{Pisani2007} and experimental observations on $sp^2$-hybridized carbon chains \cite{Thomann1985}. To emulate the charge transfer of opposite signs at the interfaces between $h$BN and ZGNR in the actual heterojunction, we set $\epsilon = \epsilon_1 > 0$ along the sites forming one edge, $\epsilon = \epsilon_2 < 0$ along the sites forming the other edge, and $\epsilon = 0$ otherwise. Our model is sketched in Figure~\ref{Fig3}(a). Further details concerning the solution of the mean-field Hubbard Hamiltonian are provided in the Supporting Information (Supporting Notes S3 and S4).

Using this model Hamiltonian, we track the evolution of the spin-dependent energy gap as a function of $\epsilon_1$ and $\epsilon_2$. This allows us to derive the electronic phase diagram shown in Figure~\ref{Fig3}(b). Depending on the magnitude of the charge residing at the edge carbon atoms in the ZGNRs, we identify three phases, all of them with zero net magnetization: (i) a spin-degenerate semiconducting phase at  $\epsilon_1 = \epsilon_2 = 0$, with finite and equal $\alpha$-spin and $\beta$-spin energy gaps; see Figure~\ref{Fig3}(c); (ii) a spin-split semiconducting phase at, e.g., $\epsilon_1 = -\epsilon_2 = 0.1t$, with a finite $\alpha$-spin energy gap that is smaller than the $\beta$-spin energy gap; see Figure~\ref{Fig3}(d); and (iii) a half-semimetallic phase at, e.g., $\epsilon_1 = -\epsilon_2 = 0.2t$, with a vanishing $\alpha$-spin energy gap and finite $\beta$-spin energy gap; see Figure~\ref{Fig3}(e).

Although we focus on the representative ZGNR of width $w=1.4$~nm, the phase diagram shown in Figure~\ref{Fig3}(b) is found to be independent of $w$. This is explained by the fact that the built-in electric field induced by boron and nitrogen atoms terminating the opposite edges of the ZGNR in the heterojuntion scales inversely with the width, and so does the energy gap of the nanoribbon [cf.\ Figure~\ref{Fig1}(c)]. Owing to this, the critical potential required to close the energy gap for one spin orientation is width-independent and is given by $(\epsilon_1 + a)^2 + (\epsilon_2 - a)^2 = b^2$, where $a = 0.19t$ and $b=0.45t$. Thus, the charge transfer from $h$BN to ZGNR is large enough to enforce the Dirac half-semimetallicity in ZGNR/$h$BN heterojunctions at any nanoribbon width [cf.\ Figure~\ref{Fig1}(f)]. Notably, the phase diagram in Figure~\ref{Fig3}(b) indicates that a half-metallic phase in the nanoribbon can appear when the  potential is exerted only on one of the two edges of the nanoribbon. However, as further corroborated by our \emph{ab initio} results shown in the Supporting Information (Supporting Note S6), a single interface between ZGNR and $h$BN is not sufficient to induce the Dirac half-semimetallic phase \cite{liu2011half}.

\begin{figure*}[t!]
    \centering
    \includegraphics[width=1.8\columnwidth]{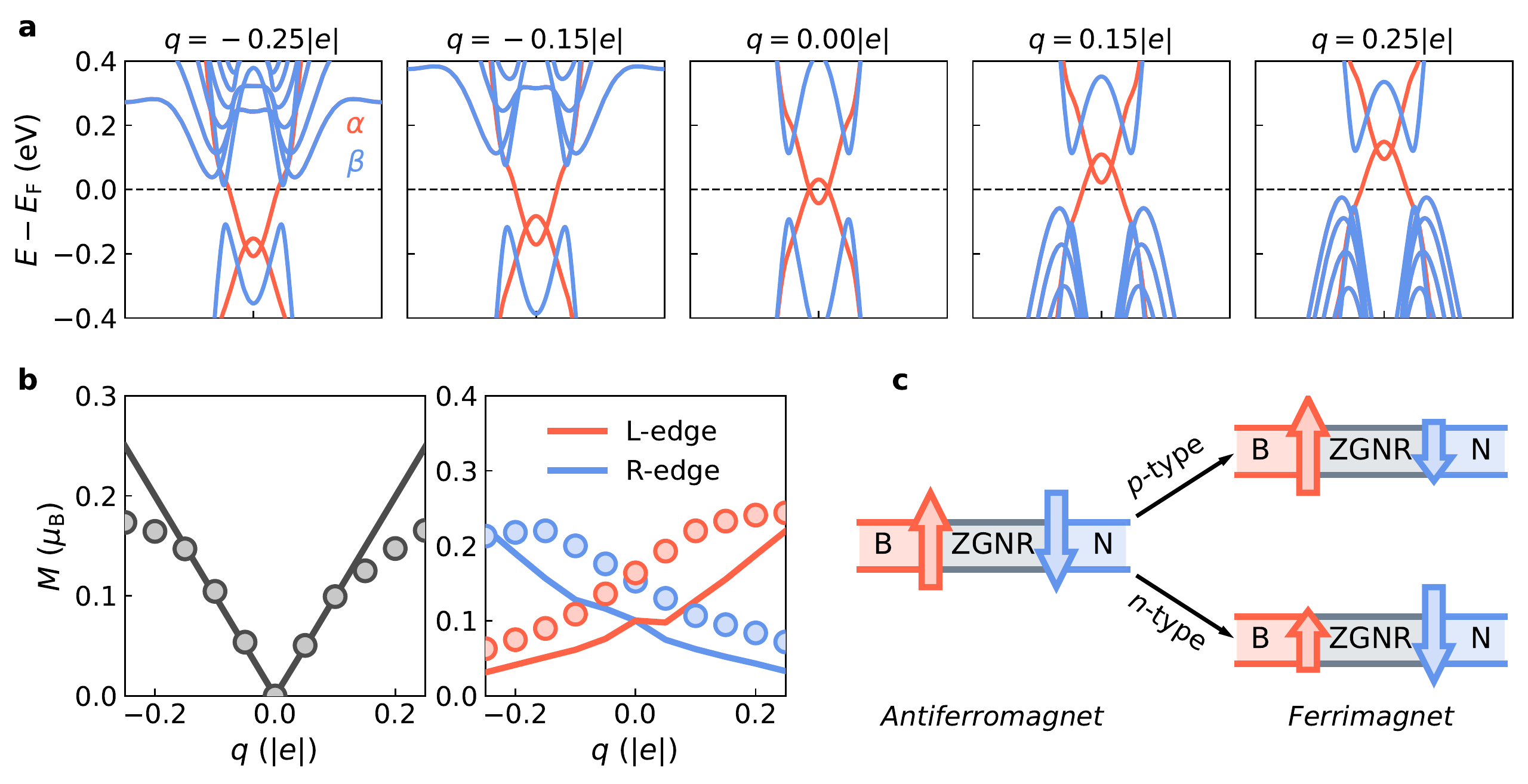}
    \caption{\textbf{Effect of charge doping on ZGNR/\textit{h}BN heterojunctions.}
    (a) Band structures of a ZGNR of width $w=1.4$~nm embedded in $h$BN for an increasing amount of excess charge ($q$) per unit cell of the nanoribbon; red and blue lines denote $\alpha$-spin and $\beta$-spin energy bands; the results are obtained at the density functional theory (DFT) level. (b) Dependence of the net magnetic moment per unit cell $M$ (left panel) and magnetic moments at the carbon atoms forming the left, boron-terminated and the right, nitrogen-terminated edge of the ZGNR (right panel), on $q$; circles denote results obtained at the DFT level, whereas solid lines denote results obtained using the mean-field Hubbard model Hamiltonian. (c) Schematics  of the doping-induced antiferromagnetic-to-ferrimagnetic phase transition in the ZGNR/$h$BN heterojunctions.}
    \label{Fig4}
\end{figure*}

As shown in Figure~\ref{Fig3}(f), the onset of the half-semimetallic phase is accompanied by an abrupt decrease of the magnetic moments on the edge atoms from $0.26\,\mu_\mathrm{B}$ to $0.07\,\mu_\mathrm{B}$, consistent with the \emph{ab initio} calculations previously discussed. This reduction of the magnetization can be understood from the electronic band diagrams in Figure~\ref{Fig2}(c): the energy gap closing results in a transfer of a fraction of $\alpha$-spin electrons to the edge occupied by $\beta$-spin electrons, thereby partially quenching the magnetic moment on the both edges.

The electronic and magnetic properties of the ZGNR/$h$BN heterojunctions can be extensively modified through charge doping. In Figure~\ref{Fig4}(a), we show the evolution of the band structure of a representative heterojunction as a function of excess charge $q$. The half-metallic character of the heterojunction is robust with respect to both $p$-type ($q > 0$) and $n$-type ($q > 0$) doping, as only $\alpha$-spin energy bands arise at the Fermi level. However, a non-trivial effect is observed in $\beta$-spin energy bands upon doping, which is not limited to a rigid energy shift. Specifically, $n$-type ($p$-type) doping  shifts selectively the $\beta$-spin conduction (valence) bands towards the Fermi level, while leaving the position of the valence (conduction) bands unchanged. This occurs because the excess charge alters only the density of $\alpha$-spin states, which affects the electron-electron repulsion experienced by $\beta$-spin states, 
but not vice versa.

Because the excess charge only affects the density of $\alpha$-spin states, doped ZGNR/$h$BN heterojunctions acquire non-vanishing net magnetic moments due to the resulting imbalance between the $\alpha$- and $\beta$-spin states, as shown in Figure~\ref{Fig4}(b). Remarkably, depending on whether the heterojunction is $n$- or $p$-doped, such net magnetic moments develop on either one or the other edge of the nanoribbon. According to the diagram of the electronic density of states given in Figure~\ref{Fig2}(c), the valence (conduction) band of the $\alpha$-spin electrons is localized on the left, boron-terminated edge (right, nitrogen-terminated edge) of the nanoribbon. Hence, $p$-type ($n$-type) doping increases the magnetic moments residing at the left (right) edge, thus enabling an electrical modulation of their spatial localization. This effect is well captured by both \emph{ab initio} and mean-field Hubbard model calculations.

Overall, as depicted in Figure~\ref{Fig4}(c), charge doping in ZGNR/$h$BN heterojunctions  drives a transition from an \textit{antiferromagnetic} ground state, where the magnetic moments at the two edges of the nanoribbon feature antiparallel orientation and equal magnitude, to a \textit{ferrimagnetic} ground state. In the latter state, the magnetic moments at the edges of the nanoribbon retain an antiparallel orientation, but their difference in magnitude results in a global spin polarization \cite{zhang2023ferrimagnets}. This is opposed to hydrogen-terminated ZGNRs, where the antiferromagnetic and spin-degenerate ground state is unchanged upon charge doping due to the spin-degeneracy of the band structure, as shown in the Supporting Information (Supporting Note S7).  Embedding ZGNRs into an $h$BN matrix is thus an effective strategy to achieve electrical control of their magnetism.

In summary, we investigated the electronic structure of recently fabricated heterojunctions consisting of zigzag graphene nanoribbons embedded in hexagonal boron nitride. We predict that these heterojunctions host a peculiar combination of Dirac half-semimetallicity and antiferromagnetism, where a fully compensated magnetization coexists with a spin-polarization of the charge carriers that arises from a Dirac semimetallic behavior in one spin orientation and a semiconducting behavior in the other. This unconventional electronic phase originates from the charge transfer of opposite polarity at the edges of the zigzag graphene nanoribbon which, in turn, leads to an energy shift of opposite signs of the otherwise spin-degenerate edge states. We have additionally shown that charge doping induces an antiferromagnetic-to-ferrimagnetic phase transition in these heterojunctions, where the net magnetic moments can be spatially modulated through the sign of the excess charge. To conclude, our findings unveil novel and highly tunable functionalities in zigzag graphene nanoribbon/hexagonal boron nitride heterojunctions emerging from an intriguing interplay between charge and spin degrees of freedom, with potential implications for integrated carbon-based spintronic devices where completely spin-polarized electrical currents are desirable.

\bigskip

M.P.\ is grateful to Daniel Bennett at Harvard University for technical assistance. N.V.T.\ acknowledges the President's PhD Scholarship of Imperial College London. E.K.\ acknowledges support from  the STC Center for Integrated Quantum Materials (NSF Grant No. DMR-1231319) 
and the Simons Foundation (Award No.\ 896626). M.P.\  is supported by the Swiss National Science Foundation (SNSF) through the Early Postdoc.Mobility Program (Grant No.\ P2ELP2-191706). We used computational resources from the National Energy Research Scientific Computing Center (NERSC), a U.S. Department of Energy Office of Science User Facility located at Lawrence Berkeley National Laboratory, operated under Contract No.\ DE-AC02-05CH11231, as well as the FASRC Cannon cluster supported by the FAS Division of Science Research Computing Group at Harvard University.

\bibliography{Final-Bibliography}

\end{document}